\documentstyle[12pt,amssymbo]{article}
\newsavebox{\uuunit}
\sbox{\uuunit}
    {\setlength{\unitlength}{0.825em}
     \begin{picture}(0.6,0.7)
        \thinlines
        \put(0,0){\line(1,0){0.5}}
        \put(0.15,0){\line(0,1){0.7}}
        \put(0.35,0){\line(0,1){0.8}}
       \multiput(0.3,0.8)(-0.04,-0.02){12}{\rule{0.5pt}{0.5pt}}
     \end {picture}}

\newcommand{\eqn}[1]{eq.(\ref{#1})}
\newcommand{\Rbar}{\Bbb{R}}

\newcommand{\ams}[1]{\Bbb{#1}}
\newcommand{\ca}[1]{{\cal #1}}
\newcommand{\unity}{\mathord{\!\usebox{\uuunit}}}

\begin{document}
\begin{titlepage}
\begin{flushright}
Preprint KUL-TF-93/22\\
hep-th/9307078
\end{flushright}

\vfill
\begin{center}
{\Large\bf The Langevin equation}\\
\vskip .5cm
{\Large\bf on a cell complex.}
\vskip 2 cm
{\bf F. Vanderseypen$^\natural$}\\
\vskip 1cm
Instituut voor Theoretische Fysica
        \\Katholieke Universiteit Leuven
        \\Celestijnenlaan 200D
        \\B-3001 Leuven, Belgium\\[0.3cm]
\end{center}
\vfill
\begin{center}
{\bf Abstract}
\end{center}
\begin{quote}
\small
We consider a cell-complex in an arbitrary Hausdorff space as a dynamical
object that can be coupled to a field defined on the complex. The Langevin
equation is then derived for this field. In other words, a noise-field is
created resulting from the field/geometry interactions.
\vspace{2mm}
\vfill
\hrule width 3.cm  {\footnotesize
\noindent $^\natural$ Aspirant N.F.W.O., Belgium,
Francois\%tf\%fys@cc3.kuleuven.ac.be}\\
\normalsize
\end{quote}
\end{titlepage}
\indent
{\large\bf Introduction.}\\
\vskip .5cm
\indent
Nobody knows how spacetime looks like at its deepest level.
But physical intuition tells us that the continuum picture is not
correct, that we use it because of its mathematical attractivity, quoting
R.Penrose : "\ldots
that to accept that there are as many points in $10^{-13}$ cm or even
$10^{-1000}$ cm as there are in the entire universe is physically
unrealistic and that our use of the continuum arises solely from its
mathematical utility\ldots"~ ~\cite{Penrose}.
We cannot deny, however, that the continuum-theories gave us deep insights
into nature and are `correct' in a classical context. So, probably the
continuum is just an approximation of a deeper structure, a structure that
`looks' like a manifold on a classical level. In the words of another
famous physicist :
"\ldots most probably, the continuity of spacetime is just an approximation
similar to the one we use in condensed matter or hydrodynamics"
\cite{Polyakov}.
One is thus led to consider physical models on non-manifold structures and
this opens the door to the various geometrical frameworks that are
available to us. Choosing one particular structure is, in my opinion,
essentially a matter of taste since there are a priori no experimental
constraints on the micro-topology of spacetime (Is it foamlike,
fuzzy,\ldots ? What is a spacetime point ?).
Still, I would like to motivate my choice for a cell
complex as a discrete background (see \cite{dis} however for some
papers that influenced this work) : \\
\\
\indent
(a) cell-complex theory is really the `differential geometry' of discrete
physical theories and as such incorporates lattice (gauge) theories, for
example : gravity on cell-complexes is equally beautiful as its continuum
counterpart~\cite{jou35}
and dimensional phase transitions appear quite naturally~\cite{jou34}.
Considering gravity on a complex with discrete values leads to
quantumgravity models~\cite{jou31}.
In other words, many physical models have a geometrical interpretation in
terms of cell-complexes.\\
\indent
(b) Simplicial decompositions,
Regge structures, etc. of manifolds have computational power
(computation of homology groups, Euler chararacteristic and Betti
numbers,\ldots). In the very same spirit, one can gain insight in geometry
of manifolds by using
cell-complexes and theories based on them can be implemented on
computers.\\
\indent
(c) One can ask what happens in the limit of
vanishing cochain volume. The nice thing about cell-complexes is the fact
that they do not only include manifolds in this limit but also fractal-like
structures \cite{jou31}. \\
\indent
(d) the Prugove\v{c}ki quantumgeometries \cite{Pru2} have a nice
generalisation
if one enlarges the space of base manifolds on which the quantumbundles are
constructed to include cell-complexes \cite{swa1}. This underlines the
versatility of cell-complexes.\\
\\
\indent
Beside this, we haven't yet consummated the fiery marriage between
spacetime and the
quantum-lady \cite{Wheeler}. Choosing a background structure has
consequences for the dynamics of the fields living on this background
space. In fact, the questions
`What is spacetime~?' and `What is a quantumfield ?' are not independent
: "\ldots `quantised fields' are not so much the product of a second
quantisation, as they are the natural manifestation of a consistently point
of view, which juxtaposes fields on quantum spacetime next to the concept
of `classical field' viewed as a field on classical
spacetime\ldots"~ ~\cite{Pru1}.
As such, quantumgeometry should describe those geometric frameworks that
can carry quantumfields and whose structure incorporates (and dictates)
quantumfield theory. In this sense it seems that cell-complex
geometry is suitable to take care of quantumfields~\cite{swa1}.
 \\
 In a previous paper~\cite{swa2} we showed that stochastic quantisation
\footnote{stochastic quantisation
\`a la
Parisi and Wu~\cite{Par}, not the formalism
elaborated by Nelson and others (\cite{Nam} and references therein)}
and stochastic geometry have the same origin, i.e. noise can be fused in
geometry such that a field defined on this geometric structure is
stochastically quantised. In this approach, however, we started from the so
called quantum noise field and derived some geometric consequences (like
the Ricci flatness of the structure in the sharp point limit). In this
paper I would like to show that one can go the other way round and that one
can actually derive the Langevin equation and the noise field starting from
the geometry of the background space (i.e. the cell-complex). Interaction
of the field with the cell-complex creates noise whose distribution is not
defined a priori but depends on the complex. I should point out that only
finite complexes are considered and that the zero (cochain) volume limit of
the distribution is not discussed.\\
\indent
A final remark. Although the cell-complex is thought here as a discrete
 spacetime model on which field theory can be elaborated, it can be used
for much more general purposes since complexes are defined in an arbitrary
Haussdorff space. So, wel will keep the formalism as general as possible.
This paper sets the stage for later developments.\\
It is organised as
follows.
We begin by presenting some generalities in the first section. Deeply
interested readers in cell-complex machinery are refered to~\cite{Cell}
(for the fancy mathematics) or~\cite{jou34,jou35} (for physical
applications).
In the next section we derive in much details the Langevin equation using
the memory function formalism. The case of a cell-complex with only
0-cochains (i.e. a set of discrete points) is then given as an example.
\\ \vskip .5cm \indent
{\bf\large Prolegomena.}\\
\vskip .5cm
\indent
This section introduces some preliminary concepts of cell-complex
theory~\cite{Cell}. We also define the cell-matter cochains which are, in the
next section, combined into a Hamiltonian to produce a dynamical system.
\\
\indent
{\sf Cell-complexes and cochains.}\\ \indent
Let $\ca{K}$ denote a finite
cell-complex of dimension $D$. The elements of
$\ca{K}$ are $p$-cells $c_p$ ($p=0,\ldots,D$) and we will assume them to be
ordered : $c_p^{i_p},\;\;p=1,\ldots,N_p$. In particular, we have a set of
ordered vertices $c_0^{i_0},\;\; i_0=1,\ldots,N_0$.
A cochain $\gamma $ on $\ca{K}$ is a function on $\ca{K}$ which is
represented by the formal sum
\begin{equation}
\gamma =\sum_{p=0}^D\sum_{i_p=1}^{N_p} \gamma (c_p^{i_p})c_p^{i_p}
\equiv \gamma^{(p)}_{i_p}\:c^{i_p}_p,
\end{equation}
where $\gamma (c_p^{i_p})$ can take values in various sets. We used the
notation $\gamma (c_p^{i_p}) \equiv\gamma^{(p)}_{i_p}$ and an implicit
summation in the last equation. We will denote by $\ca{O}(\ca{K})$ and
$\Omega (\ca{K})$
the operator-valued and real-valued cochains on $\ca{K}$ respectively.
\\
\indent
{\sf The inner product.}\\ \indent
One can define several inner products on cell-complexes, the simplest being
\begin{equation}
\langle c_p^{i_p},c_q^{i_q}\rangle= \delta _{pq}\delta ^{i_p i_q},
\label{inner}
\end{equation}
\begin{eqnarray}
\langle
f,g\rangle&=&\sum_{p,i_p}\sum_{q,i_q}(f^p_{i_p})^\dagger\;g^q_{i_q}\;
\langle c_p^{i_p},c_q^{i_q}\rangle\nonumber\\
&=& Tr(f^\dagger\;g)\nonumber\\
&=&(f^p_{i_p})^\dagger\;g^q_{i_q},\nonumber\\
\end{eqnarray}
for $f,g$ with value in $\Omega (\ca{K})$ or $\ca{O}(\ca{K})$.
One also defines
\begin{equation}
\langle f\rangle = \sum_{p,i_p} f^p_{i_p}.
\end{equation}
The space $\Omega (\ca{K})$ can be
regarded as the sum
\begin{equation}
\Omega (\ca{K}) = \Omega _0\oplus\Omega _1\oplus\ldots\oplus\Omega
_D=\bigoplus_{p=0}^D\Omega _p,
\end{equation}
where
\begin{equation}
\Omega _p=\mbox{span}\{c^{i_p}_p\mid i_p=1,\ldots,N_P\},
\end{equation}
and the subspaces $\{\Omega _i\}$ are orthogonal according to
\eqn{inner}
\begin{equation}
\langle\Omega _i,\Omega _j\rangle =0 \hskip 1.2cm \mbox{for}\;\;i\neq j.
\end{equation} \indent
{\sf The $\cdot$ and $\star$ products.}\\ \indent
Let $\ca{A}_i$ ($i=1,\ldots,D$) be the linear space generated by the
tensor products of the
$\Omega _i$'s~ ~:
\begin{equation}
\ca{A}_i = \bigoplus_{\rho =0}^\infty \left(\otimes^\rho \;\Omega
_i\right) =\bigoplus_{\rho =0}^\infty \Omega ^\rho _i,
\end{equation}
\begin{equation}
\otimes^0\Omega _j = \Omega ^0_j\equiv\Rbar\cdot\unity.
\end{equation}
Elements $\chi^{(\rho )}$ of $\Omega _i^\rho$ (cochains of rank $\rho $
and degree $i$) are thus of the form (using the condensed index notation)
\begin{eqnarray}
\chi^{(\rho )}&=&\chi^{(\rho )}_J\:c_i^J\nonumber\\
&=&\chi^{(\rho )}_{j_1\ldots j_\rho }\:c^{j_1\ldots j_\rho }_i
\hskip 1cm \rho \neq 0,\\
\chi^{(0)}&\in&\Rbar,
\end{eqnarray}
and the natural product $\cdot$ in $\ca{A}_i$ is pointwise defined :
\begin{equation}
\chi^{(\rho )}\cdot\pi^{(\sigma )} = \chi^{(\rho )}_J \:c^J_i\cdot
\pi^{(\sigma )}_K\: c^K_i=\left[ \chi^{(\rho )}_J \pi^{(\sigma
)}_K\right]\;c^J_i\,c^K_i.
\end{equation}

The space $\ca{A}_i$ with binary operations $\cdot$ and + is for
every $i=1,\ldots,D$ an algebra over
$\Rbar$. Note that $\cdot$ increases the rank
\begin{equation}
\cdot \; :
\Omega ^\rho _i\times\Omega ^\sigma _i\longrightarrow \Omega_i ^{(\rho
+\sigma)}\\
\end{equation}
It is possible to define a rank-lowering operation on $\ca{A}_i$ (the star
product) as follows :
\begin{equation}
\star\; : \left\{\begin{array}{c}
\Omega ^\rho _i\times\Omega ^\sigma _i\longrightarrow \Omega_i ^{(\rho
-\sigma)}\\
(\chi ^{(\rho )},\chi^{(\sigma )})\longmapsto \chi
^{(\rho)}\star\chi^{(\sigma )},
\end{array}\right.
\end{equation}
where we assumed that $\rho \geq\sigma$ and explicitly
\begin{equation}
\chi^{(\rho )}\star\chi^{(\sigma)} = \sum_{\{k\}} \left(\chi^{(\rho
)}_{j_1\ldots j_{\rho -\sigma }k_{1}\ldots k_{\sigma }}.
\chi^{(\sigma )}_{k_{1}\ldots k_{\sigma }}\right)c_i^{j_1}\ldots
c_i^{j_{\rho-\sigma }}.
\end{equation}
All this now can be defined equivalently for $\ca{O}(\ca{K})$ and the
star-product can be extended to include mixed products :
\begin{equation}
\star \; : \Omega _i^\rho \times\ca{O}_i^\sigma \longrightarrow
\ca{O}_i^{(\rho -\sigma )}.
\end{equation}
We note that the star product is not defined for $\rho <\sigma $ and
that $\{\ca{A}_i,\star,+\}$ is also an algebra over the real numbers.
In order to generalize $\cdot$ and $\star$ between spaces of different
degrees we set these products equal to zero for different degrees,
symbolically :
\begin{eqnarray}
\ca{A}_i\cdot\ca{A}_j&=&\delta _{ij}\:\ca{A}_i,\nonumber\\
\ca{A}_i\star\ca{A}_j&=&\delta _{ij}\:\ca{A}_i,\nonumber\\
\label{different}
\end{eqnarray}
without a summation over the index $i$.\\
\indent
{\sf The cell-matter cochains.} \\ \indent
Next we would like to have an interacting cell-matter system. To this end
we introduce operator valued cochains $\alpha $ and $\phi $ representing
the dynamics of the background (i.e. the cell-complex) and the
matter-system. Define the Fock-vacuum of the background system as
\begin{equation}
\Omega^{(b)}=
\bigoplus_{p=0}^D\bigotimes_{i_p=1}^{N_p} \Omega^{(b)}_{(p,i_{p})},
\end{equation}
where the object $\Omega^{(b)}_{(p,i_{p})}$ is the vacuum state of the
$i_p$-th $p$-cell satisfying the relation (no summation)
\begin{equation}
a^{(p)}_{i_p}\Omega^{(b)}_{(p,i_{p})}=0.
\end{equation}
The operators $a^{(p)}_{i_p}$ are defined by the commutation relations :
\begin{equation}
[a^{(p)}_{i_p},(a^{(p^\prime)}_{i_{p^\prime}})^\dagger]=\delta
_{i_{p^\prime}i_p}\delta ^{p^\prime p},
\end{equation}
\begin{equation}
[a^{(p)}_{i_p},a^{(p^\prime)}_{i_{p^\prime}}]=
[(a^{(p)}_{i_p})^\dagger,(a^{(p^\prime)}_{i_{p^\prime}})^\dagger]= 0.
\end{equation}
In the very same way we define
\begin{equation}
\Omega ^{(s)}=\bigoplus_{p=0}^D\bigotimes_{i_p=1}^{N_p} \Omega
^{(s)}_{(p,i_p)},
\end{equation}
\begin{equation}
A^{(p)}_{i_p}\Omega^{(s)}_{(p,i_p)}=0,
\end{equation}
\begin{equation}
[A^{(p)}_{i_p},(A^{(p^\prime)}_{i_{p^\prime}})^\dagger]=\delta
_{i_{p^\prime}i_p}\delta ^{p^\prime p},
\end{equation}
\begin{equation}
[A^{(p)}_{i_p},A^{(p^\prime)}_{i_{p^\prime}}]=
[(A^{(p)}_{i_p})^\dagger,(A^{(p^\prime)}_{i_{p^\prime}})^\dagger]= 0,
\end{equation}
for the matter system $\phi $ living on the cell-complex.
The total Fock-vacuum of the interacting cell-matter system is, of course,
\begin{equation}
\Omega ^{(t)}=\Omega ^{(b)}\otimes\Omega ^{(s)}.
\end{equation}
Now we are ready
to
define the $\alpha $ and $\phi $ cochains. They have the following
expressions
\begin{equation}
\alpha =\sum_p\sum_{i_p}
\left[\oplus_{q}\otimes_{i_q} \alpha _{(p,q)}\right]\;\, c_p^{i_p},
\end{equation}
\begin{equation}
\phi =\sum_p\sum_{i_p}
\left[\oplus_{q}\otimes_{i_q} \phi_{(p,q)}\right]\;\, c_p^{i_p}.
\end{equation}
The coefficients $\alpha _{(p,q)}$ and $\phi _{(p,q)}$ of the cochain
expansions
are operators acting on the Fock vacua defined above, they are defined as
\begin{equation}
\alpha _{p,q} =\left\{\begin{array}{ll}
\unity & p\neq q\\
a^{(p)}_{i_p} & p=q,
\end{array}\right.
\end{equation}
\begin{equation}
\phi_{p,q} =\left\{\begin{array}{ll}
\unity & p\neq q\\
A^{(p)}_{i_p} & p=q.
\end{array}\right.
\end{equation}
But to simplify the notation we will abreviate all this to :
\begin{equation}
\alpha =
a^{(p)}_{i_p} c_p^{i_p},
\end{equation}
and to
\begin{equation}
\phi =
A^{(p)}_{i_p} c_p^{i_p}.
\end{equation}
\\
\vskip .5cm \indent
{\bf\large Derivation of the Langevin equation.}\\
\vskip .5cm
\indent
The derivation of the generalised Langevin equation on the cell-complex is
done in several steps. First we combine $\phi $ and $\alpha $, introduced
above, into a ket-cochain $\mid\!\psi\rangle$ whose coeffiecients are
operators on the Fock-space and as such depends on the choosen Fock-space
state vector. Formally this then means that in the derivation, given below,
we assume that a certain Fock-state has been picked out a priori. The
fundamental ingredient in the derivation is the Hamiltonian \eqn{Hamil}
depending on a $2\times 2$~-~matrix~ ~$\Gamma$~ ~:
\begin{equation}
\Gamma =
\left(\begin{array}{cc}
\omega ^{(b)} & \omega ^{(i)}\\
\tilde\omega ^{(i)}& \omega ^{(s)}\\
\end{array}\right),
\end{equation}
where $\omega ^{(b)}$ describes the dynamics of the cell-complex, $\omega
^{(s)}$ that of the field-system and $\omega ^{(i)}$, $\tilde\omega ^{(i)}$
give the interaction terms in $\ca{H}$. To solve the evolution equation~
{}~\eqn{VN} we use the Laplace transformation and projectors. The resulting
equation \eqn{totale} is then transformed back in $t$-space and gives us
the (generalised) Langevin equation \eqn{Langekort}. The simplest model on
which we can apply \eqn{Langekort} is in the case of a zero cell-complex,
i.e. a set of points. This is done at the end of this section.
\\
\indent
{\sf The bra-ket notation.}\\ \indent
We combine the the background system and the field system into a
vector-like object~ ~:
\begin{eqnarray}
\psi &=& \left(\begin{array}{c}
\alpha \\ \phi
\end{array}\right)\nonumber\\
&=& \left(\begin{array}{c}
a^{(p)}_{i_p}\:c^{i_p}_{p}\\ A^{(p)}_{i_p}\:c^{i_p}_{p}
\end{array}\right)
= \left(\begin{array}{c}
a^{(p)}_{i_p}\\ A^{(p)}_{i_p}
\end{array}\right)\;c^{i_p}_{p}.\nonumber\\
\end{eqnarray}
In applying the memory-function formalism we will use a bra-ket notation,
\begin{eqnarray}
\mid\!\psi\rangle &=&\left(\begin{array}{c}
a^{(p)}_{i_p}\\ A^{(p)}_{i_p}
\end{array}\right)\mid c^{i_p}_{(p)}\rangle \nonumber\\
&=& \psi^{(p)}_{i_p}\mid c^{i_p}_{p}\rangle,
\end{eqnarray}
\begin{equation}
\langle c_p^{i_p}\mid c_q^{i_q}\rangle\equiv \langle c_p^{i_p},
c_q^{i_q}\rangle= \delta _{pq}\delta ^{i_p i_q}.
\end{equation}
Remenber that a sumation is implied over repeated indices.
\\ \indent
{\sf The Hamiltonian.}\\
\indent
Assume now that $\Gamma $ is a two-by-two (real) matrix whose entries are
elements of $\oplus_p\Omega^2_p$~ ~:
\begin{eqnarray}
\Gamma &=&\left(\Gamma ^{\mu \nu }\right)\hskip 1.3cm \mu,\nu=1,2
\nonumber\\
\Gamma ^{\mu \nu }&\in& \oplus_p\Omega^2_p.\nonumber\\
\end{eqnarray}
We set the Hamiltonian of the interacting system equal to
\begin{equation}
\ca{H} = \langle \psi,\Gamma \star\psi\rangle\equiv\langle \psi\mid\Gamma
\mid\!\psi\rangle.
\label{Hamil}
\end{equation}
In component form this reads as
\begin{eqnarray}
\ca{H}&=&\langle \psi\mid\Gamma \mid\!\psi\rangle\nonumber\\
&=& (\psi^{(p)}_{i_p})^\dagger\:(\Gamma \star\psi)^{(p)}_{i_p}\nonumber\\
&=& (\psi^{(p)}_{i_p})^\dagger\:\Gamma
^{(p)}_{i_pi_k}\:\psi^{(p)}_{i_k},\\
\end{eqnarray}
and in this final form there is still an implicit matrix multiplication.
The dynamics is found by solving the Von Neumann equation
\begin{equation}
\frac{d\psi(t)}{dt} = \imath [\ca{H},\psi(t)] = -\imath\Gamma \star\psi(t),
\label{VN}
\end{equation}
which is easily solved in terms of the Liouvillian $\Gamma $ :
\begin{eqnarray}
\langle\psi(t)\!\mid &=& \langle \psi\mid\exp\imath\:t\Gamma,\nonumber\\
\mid\!\psi\rangle&\equiv&\mid\psi(0)\rangle.\nonumber\\
\label{Liouville}
\end{eqnarray}
We define the Laplace transform for a $t$-dependent function $q$ as :
\begin{equation}
\mbox{Lap}(q)=q(z) = \int_0^\infty dt e^{\imath zt} q(t),
\end{equation}
and applying this transformation to \eqn{Liouville} results in
\begin{eqnarray}
\mbox{Lap}(\langle\psi\!\mid)&=&\int_0^\infty dt e^{-\imath
zt}\langle\psi(t)\!\mid\nonumber\\
&=& \langle\psi\!\mid\frac{\imath}{z-\Gamma }\nonumber\\
&=&\langle\psi(z)\!\mid.\nonumber\\
\label{psiz}
\end{eqnarray}
\indent
{\sf Projectors and memory functions.}\\ \indent
 We transform the time evolution equation \eqn{Liouville} into a
differential equation using the memory-function formalism~ ~\cite{Mori}.
First define the projection operators $\ams{P}$ and $\ams{Q}$ as
\begin{equation}
\ams{P}+\ams{Q} = \unity,
\end{equation}
\begin{equation}
\ams{P} =C^{-1}_0\, \mid\!\psi\rangle\langle\psi\!\mid,
\end{equation}
\begin{equation}
C_0\equiv\langle\psi\!\mid\!\psi\rangle.
\end{equation}
Inserting the trivial identity $\Gamma =\Gamma \ams{Q}+\Gamma
\ams{P}$ in \eqn{psiz} and using the operator-identity
\begin{equation}
\frac{1}{A+B}=\frac{1}{A}-\frac{1}{A}\: B\:\frac{1}{A+B}
\end{equation}
gives
\begin{equation}
\langle\psi(z)\!\mid=\langle\psi\!\mid\left\{\frac{\imath}{z-\Gamma
\ams{Q}}+\frac{1}{z-\Gamma \ams{Q}}\Gamma \ams{P}\frac{\imath}{z-\Gamma
}\right\}.
\label{psiexp}
\end{equation}
The first term can be transformed into a more esthetic
form :\\
first we have that
\begin{equation}
\frac{\imath}{z-\Gamma \ams{Q}}= \frac{\imath}{z}\:\left\{1+\frac{
\Gamma \ams{Q}}{z-\Gamma \ams{Q}}\right\}\nonumber\\
\end{equation}
and multiplying the last term of this equality from the left with
$\langle\psi\!\mid$ :
\begin{eqnarray}
\imath\langle\psi\!\mid\frac{ \Gamma \ams{Q}}{z-\Gamma \ams{Q}}&=&
\langle\dot\psi\!\mid\frac{\ams{Q}}{z-\Gamma \ams{Q}}\nonumber\\
&=&\langle\dot\psi\!\mid\ams{Q}\frac{1}{z-\ams{Q}\Gamma}\ams{Q},\nonumber\\
&=& \langle\dot\psi\!\mid\ams{Q}\frac{1}{z-\ams{Q}\Gamma
\ams{Q}}\ams{Q}\nonumber\\
\end{eqnarray}
giving altogether
\begin{equation}
z\:\langle\psi\!\mid\frac{\imath}{z-\Gamma
\ams{Q}}=\imath\langle\psi\!\mid+\langle\dot\psi\!\mid
\ams{Q}\frac{1}{z-\ams{Q}\Gamma \ams{Q}}\ams{Q}.
\label{first}
\end{equation}
In this derivation we used the notation $\langle\dot\psi\!\mid$ for
$\langle\dot\psi(0)\!\mid=\imath\langle\psi\!\mid\Gamma $ and the nilpotenty of
$\ams{Q}$.
The second part of \eqn{psiexp}, on the other hand, is equivalent
to \begin{eqnarray}
z\:\langle\psi\!\mid\frac{1}{z-\Gamma \ams{Q}}\Gamma
\ams{P}\frac{\imath}{z-\Gamma
}&=& \langle\psi\!\mid\frac{1}{z-\Gamma \ams{Q}}\Gamma
\mid\!\psi\rangle C^{-1}_0\langle\psi\!\mid\frac{\imath}{z-\Gamma
\ams{Q}}\nonumber\\
&=&\left(\Omega -\imath\Sigma (z)\right)\langle\psi(z)\!\mid,\nonumber\\
\label{sec}
\end{eqnarray}
where we introduced the so-called memory functions $\Omega $ and $\Sigma
(z)$ defined as
\begin{equation}
\Omega =\langle\psi\!\mid\Gamma \mid\!\psi\rangle\,C^{-1}_0,
\end{equation}
\begin{equation}
\Sigma (z) = \imath\langle\dot\psi\!\mid\ams{Q}\frac{1}{z-\Gamma
\ams{Q}}\mid\dot\psi\rangle\,C^{-1}_0.
\end{equation}
finally glueing \eqn{first} and \eqn{sec} together gives :
\begin{equation}
(z-\Omega +\imath\Sigma (z))\langle\psi(z)\!\mid= \imath\langle\psi\!\mid+
\langle\dot\psi\!\mid
\ams{Q}\frac{1}{z-\ams{Q}\Gamma \ams{Q}}\ams{Q}.
\label{totale}
\end{equation}
\indent
{\sf The Langevin equation.}\\ \indent
Define the correlator $C(z)$ as
\begin{equation}
C(z)=\langle\psi\!\mid\frac{\imath}{z-\Gamma
}=\langle\psi(z)\!\mid\!\psi\rangle,
\end{equation}
because $\ams{Q}\mid\!\psi\rangle=0$ we get
\begin{equation}
C(z) = \frac{i C_0}{z-\Omega +\imath\Sigma (z)}.
\label{Cz}
\end{equation}
Let us also introduce the time-equivalent $\sigma (t)$ of the memory
function :
\begin{equation}
\Sigma (z)= \mbox{Lap}(\sigma ).
\end{equation}
After some algebraic manipulations, one easily gets the explicit form for
$\sigma (t)$
\begin{equation}
\sigma (t) =\langle\dot\psi\!\mid\ams{Q} e^{-\imath\,t\, \ams{Q}\Gamma
\ams{Q}} \ams{Q}\mid\dot\psi\rangle\,C^{-1}_0,
\end{equation}
and finally, if we apply the inverse Laplace transform to \eqn{Cz}
and to \eqn{totale} we get
the  promised Langevin differential equation :
\begin{equation}
(\partial _t +\imath\Omega )C(t) +\int_0^t d\tau\sigma (t-\tau )C(\tau ) =
0\hskip 1cm t>0,
\label{Langekort}
\end{equation}
and
\begin{equation}
(\partial _t +\imath\Omega )\langle\psi(t)\!\mid +\int_0^t d\tau\sigma
(t-\tau )\langle\psi(\tau )\mid =
\imath\langle\eta(t)\mid\hskip 1cm t>0,
\label{Languit}
\end{equation}
with the $\eta$-field (the so-called generalised noise field) defined as :
\begin{equation}
\langle\eta(t)\mid = \langle\dot\psi\!\mid\ams{Q}
e^{-\imath\,t\, \ams{Q}\Gamma \ams{Q}} \ams{Q}.
\end{equation}
Since $\ams{Q}\mid\!\psi\rangle =0$ we have $\langle\eta(t)\mid\!\psi\rangle=0$
and also
\begin{equation}
\langle\eta(t)\mid\eta(t^\prime)\rangle=\langle\dot\psi\!\mid\ams{Q}\:\exp[-
\imath\,(t-t^\prime)\ams{Q}\Gamma \ams{Q}]\:\ams{Q}\mid\dot\psi\rangle.
\end{equation}
So, \eqn{Languit} implies \eqn{Langekort}.
\vskip .5cm
Let us take the following simple case.
Assume that
$\Gamma $ has the following form :
\begin{equation}
\left(\begin{array}{cc}
\omega ^{(b)} & \unity\\
\unity & \omega ^{(s)}\\
\end{array}\right),
\end{equation}
with $\omega ^{(b)}$ equal to
\begin{equation}
\omega ^{(b)} = \sum_{kl} \delta_{kl}\omega _k c^k_0c^l_0\in \Omega _0^2,
\end{equation}
while $\omega ^{(s)}$ is an arbitrary element of $\Omega^2_0$.
Now, it is easily seen that for the ground state of the total space
$\Omega ^{(t)}$ we have
\begin{eqnarray}
\ams{P}&=& \mid\!\psi\rangle\langle\psi\!\mid\nonumber\\
&=&[\Omega ^{(t)}]^\dagger\;\left(\begin{array}{cc}
a_k\,a^\dagger_l& a_k\,A^\dagger_l\\
A_k\,a^\dagger_l& A_k\,A_l^\dagger\\
\end{array}\right)\;[\Omega ^{(t)}]\:\mid c_0^k\rangle\langle
c_0^l\mid\nonumber\\
&=& \left(\begin{array}{cc}
\delta _{kl}& 0\\
0&\delta _{kl}
\end{array}\right)
\:\mid c_0^k\rangle\langle c_0^l\mid\nonumber\\
&=&\unity.
\label{unity}
\end{eqnarray}
This implies that $\ams{Q}=0$ and \eqn{Languit} reduces to
\begin{equation}
(\partial _t +\imath\Omega )\langle\psi(t)\!\mid=0 \hskip 1.2cm t>0.
\label{Langex}
\end{equation}
The dynamical equations read then explicitly
\begin{equation}
\frac{da_k(t)}{dt}= \imath \:\omega _ka_k(t) +\imath \:A_k(t),
\label{kleinea}
\end{equation}
and
\begin{equation}
\frac{dA_k(t)}{dt} = \imath\: \sum_l \omega ^{(s)}_{kl} A_l(t)+\imath\:
a_k(t), \label{grotea}
\end{equation}
as can be checked directly from the Hamiltonian. Of course
\eqn{Langex} could have been derived directly from \eqn{Liouville}
by differentiating it with respect to $t$ and putting \eqn{unity} in
sandwich between $\Gamma $ and the exponential. Note also that \eqn{unity}
implies $\ca{H} = \Omega $.
Solving \eqn{kleinea} and substituting this into \eqn{grotea} leads to the
following cochain equation~ ~:
\begin{equation}
\frac{d\phi(t)}{dt}= i\,\omega ^{(s)}\star\phi
(t)+\imath\xi^\dagger(t)-M_\phi(t), \end{equation}
\begin{eqnarray}
M_\phi (t) &=& \int_0^td\tau e^{-i\omega _k(\tau -t)}\,A_k(\tau )\;c_0^k,\\
\xi(t) &=& e^{-\imath\omega _k t}a_k^\dagger(0)\:c_0^k.
\end{eqnarray}
Taking the inner product of $\xi$ and $\xi^\dagger$ gives
\begin{eqnarray}
\langle\xi(t)\rangle&=& \sum_k e^{-\imath\omega _k t}a_k^\dagger(0)\\
\langle \xi
(t^\prime),\xi(t)\rangle&=&Tr\left(\xi^\dagger
(t^\prime)\xi(t)\right)\nonumber\\
&=&\sum_k \left[\xi^\dagger(t^\prime)\right]_k
\left[\xi(t)\right]_k \nonumber\\
&=& \sum_k e^{\imath\omega
_k(t^\prime-t)}\,a_k(0)a^\dagger_k(0).\nonumber\\ \end{eqnarray}
In the ground state $\Omega ^{(t)}$ we thus have
\begin{equation}
\left[\langle\xi(t)\rangle\right]_{\Omega ^{(t)}}=0,
\end{equation}
and
\begin{equation}
\left[\langle \xi
(t^\prime),\xi(t)\rangle\right]_{\Omega^{(t)}}= \sum_k
e^{\imath\omega _k(t^\prime-t)}.
\label{corr}
\end{equation}
So we see that $\xi$ represents a noise term with correlations as described
by \eqn{corr}, while $M_\phi $ represents the effect of the field at
earlier times on itself (memory effect). It is tempting to replace the
finite sum of the 2-point correlation function by a $\delta $-function in
the zero cochain-volume, but this needs a rigorous proof.
\\ \vskip .5cm \indent
{\bf\large Conclusion.}\\
\vskip .5cm
\indent
We described a dynamical system on a cell-complex leading to the Langevin
equation. It was shown that a (generalised) noise field is produced as an
integral part of the dynamics. The main restriction on the interaction is
\eqn{different}, since this does not allow interactions between cochains of
different degrees. We didn't discuss the continuum limit, which is a
profound (mathematical) problem.

\vskip 1.4cm

\begin{center}
{\tt I wish to thank K.Siemion, B.Dezillie, F.De Jonghe,\\
A.K.Callebaut, B.Meeus\\
and especially C.Detroye for their patience and support.}\\
\end{center}

\end{document}